# The Interstellar Medium In Galaxies Seen A Billion Years After The Big Bang

P.L. Capak[1,2*], C. Carilli[3,4], G. Jones[5], C. M. Casey[6], D. Riechers[7], K. Sheth[8], C. M. Carollo[9], O. Ilbert[10], A. Karim[11], O. LeFevre[10], S. Lilly[8], N. Scoville[2], V. Smolcic[12], L. Yan[1,2]

[1]Infrared Processing and Analysis Center (IPAC), 1200 E. California Blvd., Pasadena, CA, 91125, USA E-mail: capak@ipac.caltech.edu
[2]California Institute of Technology, 1200 E. California Blvd., Pasadena, CA, 91125, USA
[3]National Radio Astronomy Observatory, P.O. Box 0, Socorro, NM 87801, USA
[4]Astrophysics Group, Cavendish Laboratory, J. J. Thomson Avenue, Cambridge CB3 0HE, UK
[5]New Mexico Institute of Mining and Technology, 801 Leroy Pl, Socorro, NM 87801, USA
[6]Department of Astronomy, The University of Texas at Austin, 2515 Speedway, Stop C1400, Austin, Texas 78712, USA
[7]Department of Astronomy, Cornell University, 220 Space Sciences Building, Ithaca, NY 14853, USA
[8]National Radio Astronomy Observatory, 520 Edgemont Road, Charlottesville, VA 22903, USA
[9]Institute for Astronomy, ETH Zurich, CH-8093 Zurich, Switzerland
[10]Aix Marseille Université, CNRS, LAM (Laboratoire d'Astrophysique de Marseille), UMR 7326, 13388, Marseille, France
[11]Argelander-Institut für Astronomie, Auf dem Hügel 71, D-53121 Bonn, Germany
[12]Physics Department, University of Zagreb, Bijenička cesta 32, 10002 Zagreb, Croatia

**Evolution in the measured rest frame ultraviolet spectral slope and ultraviolet to optical flux ratios indicate a rapid evolution in the dust obscuration of galaxies during the first 3 billion years of cosmic time (z>4)[1-3]. This evolution implies a change in the average interstellar medium properties, but the measurements are systematically uncertain due to untested assumptions[4,5], and the inability to measure heavily obscured regions of the galaxies. Previous attempts to directly measure the interstellar medium in normal galaxies at these redshifts have failed for a number of reasons[6-9] with one notable exception[10]. Here we report measurements of the [CII] gas and dust emission in 9 typical (~1-4L*) star-forming galaxies ~1 billon years after the big bang (z~5-6). We find these galaxies have >12x less thermal emission compared with similar systems ~2 billion years later, and enhanced [CII] emission relative to the far-infrared continuum, confirming a strong evolution in the interstellar medium properties in the early universe. The gas is distributed over scales of 1-8 kpc, and shows diverse dynamics within the sample. These results are consistent with early galaxies having significantly less dust than typical galaxies seen at z<3 and being comparable to local low-metallicity systems[11].**

We have obtained rest-frame far-infrared (FIR) (3-1000µm) measurements of 9 "normal" (~1-4 L*) galaxies and one low luminosity quasar redshifts of z=5-6, ~1 billion years after the big bang. These data were taken with an early 20-antenna version of Atacama Large Millimeter Array (ALMA) in band 7 (1090-800µm) continuum mode with 7.5GHz of bandwidth, ~20 minutes of integration, and with the compact array yielding a resolution of ~0.6 arc seconds. One of the four side bands was always centered on the 158µm [CII] line, which is the dominant FIR cooling line for neutral gas in normal star forming galaxies[12,13] and hence is a good indicator of galaxy dynamics and the spatial extent of the ISM[13,14]. The other three spectral bands yield a measurement of the dust continuum emission at $\lambda_o$~150µm, which is a good indicator of the total infrared luminosity for normal FIR spectral energy distributions (SEDs) (Method 4).

The sample of objects was selected as Lyman Break Galaxies (LBGs) from the 2 square degree Cosmic Evolution Survey (COSMOS) field[15] and have spectroscopically determined absorption line redshifts from the Deep Extragalactic Imaging Multi-Object Spectrograph (DEIMOS) on the W. M. Keck-II Observatory in Hawaii. The objects were selected to be "characteristic" with luminosities between 1 and 4 $L_*$ and with UV spectral slopes (β) between -1.4 and -0.7. Three objects (1, 2, and 10) also meet the selection criteria for Ly-α emitters at z~5.7. When possible, UV morphologies are measured from the Hubble-Space-Telescope (HST) Advanced Camera For Surveys (ACS) F814W (0.8µm) data[16]. Stellar masses were determined by fitting SED models to existing COSMOS data and new 3-5µm photometry from the Spitzer-SPLASH survey[17] (Method 3). Rest frame infrared luminosities were determined by assuming the range of far infrared properties measured at z~2-5[18] and have a systematic uncertainty of 0.3 dex due to the uncertainty in the shape of the FIR SED (Method 4).

We detect 4 of 9 galaxies in the dust continuum (Figure 1, Extended Data Table 1), and a mean combination (stacking) the data on the remaining six objects yields a continuum detection of 35±13µJy. This is more than 12x lower than expected for systems with similar UV properties at z<3 assuming the worst-case systematics in our FIR determination (Method 5). This can be quantified with the Infrared-Excess to UV-slope (IRX-β) relation[4] (Figure 2), which relates the amount of dust absorption measured in the UV to the amount of infrared re-emission and is sensitive to both dust properties and column density. We find our continuum detected objects are consistent with the Small Magellanic cloud (SMC)[19], which has lower metallicity gas, and less thermal FIR emission[20] than typical galaxies at z<3. Our undetected systems have even less thermal emission than the SMC. Hence, we conclude z~6 galaxies are significantly less dust obscured than more evolved systems at z<3, confirming previous results[1-3,6].

The very low IRX values at a β ~ -1.2 implied by our non-detections are difficult to explain in the context of the very young systems expected at z=5-6[21]. The simplest explanation is going to the extremes of the 1σ systematics in our estimates of β and LIR, which would imply relatively un-obscured systems with hot dust (Method 4). But, we cannot exclude changes in the dust geometry or rapid disruption of the molecular clouds that could lead to lower IRX vales at a given β with more normal dust temperatures[22]. Higher signal-to-noise near-infrared photometry and shorter wavelength FIR data will ultimately be needed to understand these sources.

A corollary of this result is that UV derived star-formation rates at high redshift (z>5) should use the SMC like IRX-β relation or assume no dust rather than the currently assumed Meurer IRX-β relation. This will decrease the UV derived star-formation rates by a typical factor of ~2-4 for individual galaxies from those implied by the Meurer relation for similar values of β. The effect on the global star-formation history is much smaller, <40%, because the majority of star-formation is in low luminosity (<L*) galaxies that were already assumed to have little or no dust extinction (Method 8, Extended Data Figure 4).

In contrast to the dust emission, we find >3σ detections of the [CII] line in all 9 normal galaxies (Figure 3). The line emission is spectrally resolved in all cases, with [CII] velocity dispersions of σ=63-163 km/s and marginally resolved at our spatial resolution of ~0.5-0.9" indicating galaxies with $M_{dyn}$~$10^{9-11} M_\alpha$ (Method 2, 4). We also detect two optically faint [CII] emitters at redshifts consistent with the targeted objects. HZ5a is detected near HZ5 at a redshift consistent with the in-falling gas seen in the optical spectra of HZ5[23]. HZ8W corresponds to an optically faint companion to HZ8 and has a similar redshift. Taken together the direct and serendipitous detections suggest ubiquitous and enhanced [CII] emission in early galaxies similar to that seen in local low-metallicity systems[11] (Figure 4).

The [CII] enhancement in local systems is caused by a lower dust to gas ratio which allows the UV radiation field to penetrate a larger volume of molecular cloud[11]. Our significantly lower IRX values and enhanced [CII]/FIR ratios would suggest a similar effect is happening in high redshift galaxies. But, evolution in metal abundances which change the intrinsic UV slope, changes in the dust properties, and differences in the dust geometry, have also been suggested as possible causes of the transition in obscuration properties with redshift[24]. The systems in this study were selected to have broad UV absorption features in their spectra that indicate a relatively homogeneous metal abundance of ≈0.25 solar. At this metallicity the UV spectral slopes are expected to be similar to those of solar metallicity systems[25]. Furthermore, the population these objects were selected from have dust attenuation properties similar to lower redshift objects[26]. So changes in the dust properties that would be measureable in the attenuation curve are likely not the primary factor. At fixed stellar mass, our sample has physical properties (gas velocity dispersions, sizes, and star-formation rates) that imply geometries consistent with z~1-3 galaxies that exhibit dust properties like those found in the Meurer model[27]. So, changes in the average cloud geometry are not obvious, but we cannot fully exclude this possibility since it would explain the observed IRX properties. So we conclude, a significant decrease in the dust to gas ratio is the most likely explanation for the evolution in extinction properties.

These results are seemingly at odds with previous measurements[6-9] that failed to detect [CII] emission at high redshift. Our sample contains objects with both Ly-α emission and absorption, and the [CII] emission strength is uncorrelated with the Ly-α properties (Extended Data Table 1), which means the use of Ly-α redshifts is not the cause. The most likely explanation for the lack of detections is observational uncertainty in the expected line flux, line width, and frequency. Even obscuration uncorrected UV star formation rates will over-estimate the expected FIR continuum flux because less than 20% of the star formation is leading to FIR emission in our sample[6] (Figure 2). This means the FIR emission of galaxies will be more difficult to detect than previously expected based on z<3 scaling relations (Method 5,6) so longer integration times are required. That said, these data show [CII] is enhanced and readily detectable with ALMA at z~6 if moderate integration times are used. Finally, the [CII] lines are resolved in velocity space so single channel sensitivity should not be used for limits (Method 6).

**Acknowledgements:** Support for this work was provided by NASA through an award issued by JPL/Caltech. We would like to thank the ALMA staff for facilitating the observations and aiding in the calibration and reduction process. ALMA is a partnership of ESO (representing its member states), NSF (USA) and NINS (Japan), together with NRC (Canada) and NSC and ASIAA (Taiwan), in cooperation with the Republic of Chile. The Joint ALMA Observatory is operated by ESO, AUI/NRAO and NAOJ. This work is based in part on observations made with the Spitzer Space Telescope, and the W.M. Keck Observatory along with Archival data from the NASA/ESA Hubble Space Telescope, the Subaru Telescope, the Canada-France-Hawaii-Telescope, the ESO Vista telescope obtained from the NASA/ IPAC Infrared Science Archive. VS acknowledges funding by the European Union's Seventh Frame-work program under grant agreement 337595 (ERC Starting Grant, 'CoSMass').

**Author Information:** Please send correspondence to Peter Capak, capak@astro.caltech.edu. This paper makes use of ALMA data: ADS/JAO.ALMA#2012.1.00523.S. ALMA

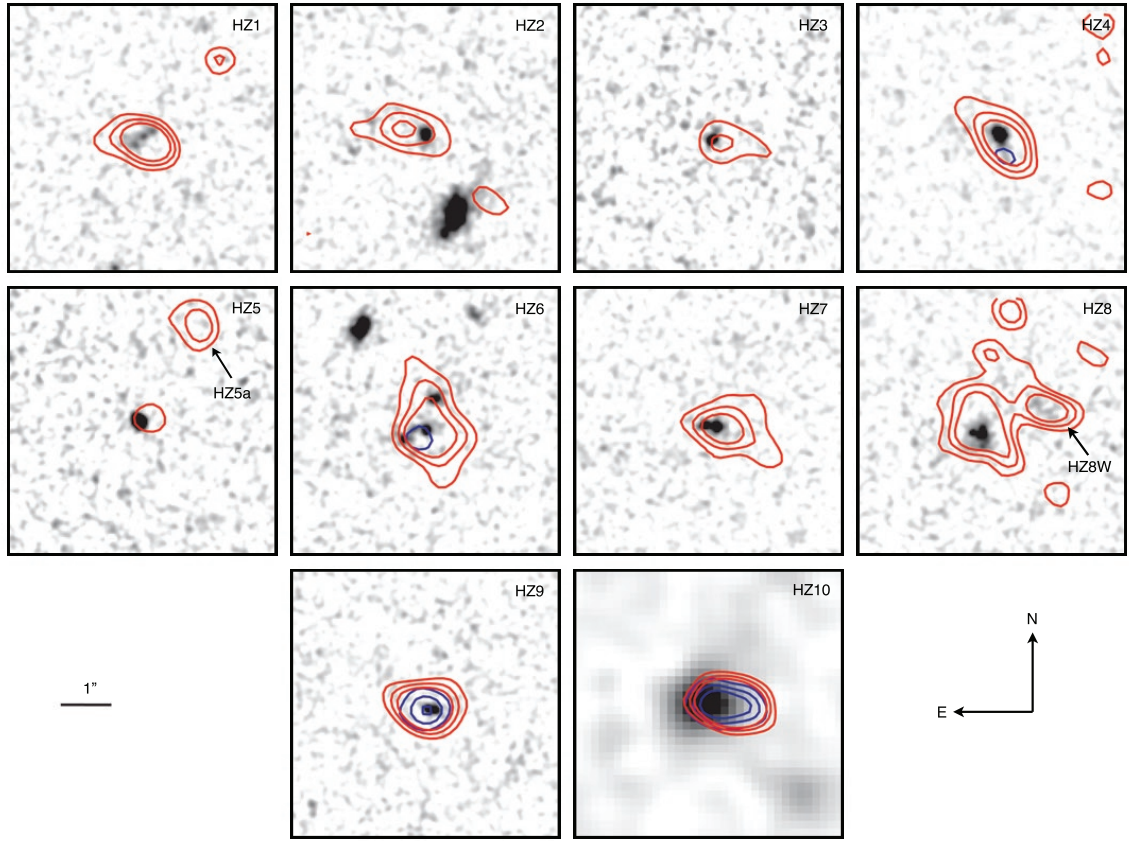

**Figure 1 |** The [CII] line detections (red contours) and weak ~158μm FIR continuum detections (blue contours) are shown with the rest frame UV images as the background. The images are 5"x5" and the contours are 2, 6 and 10 σ with [CII] line profiles for each source shown in Figure 3. The background images are from HST-ACS in the F814W[16] band where the morphologies will be affected by Ly-α, except for HZ10, which is Subaru z' band. All objects are detected in [CII] showing that a large amount of gas is present in these systems, but only 4 are detected in continuum.

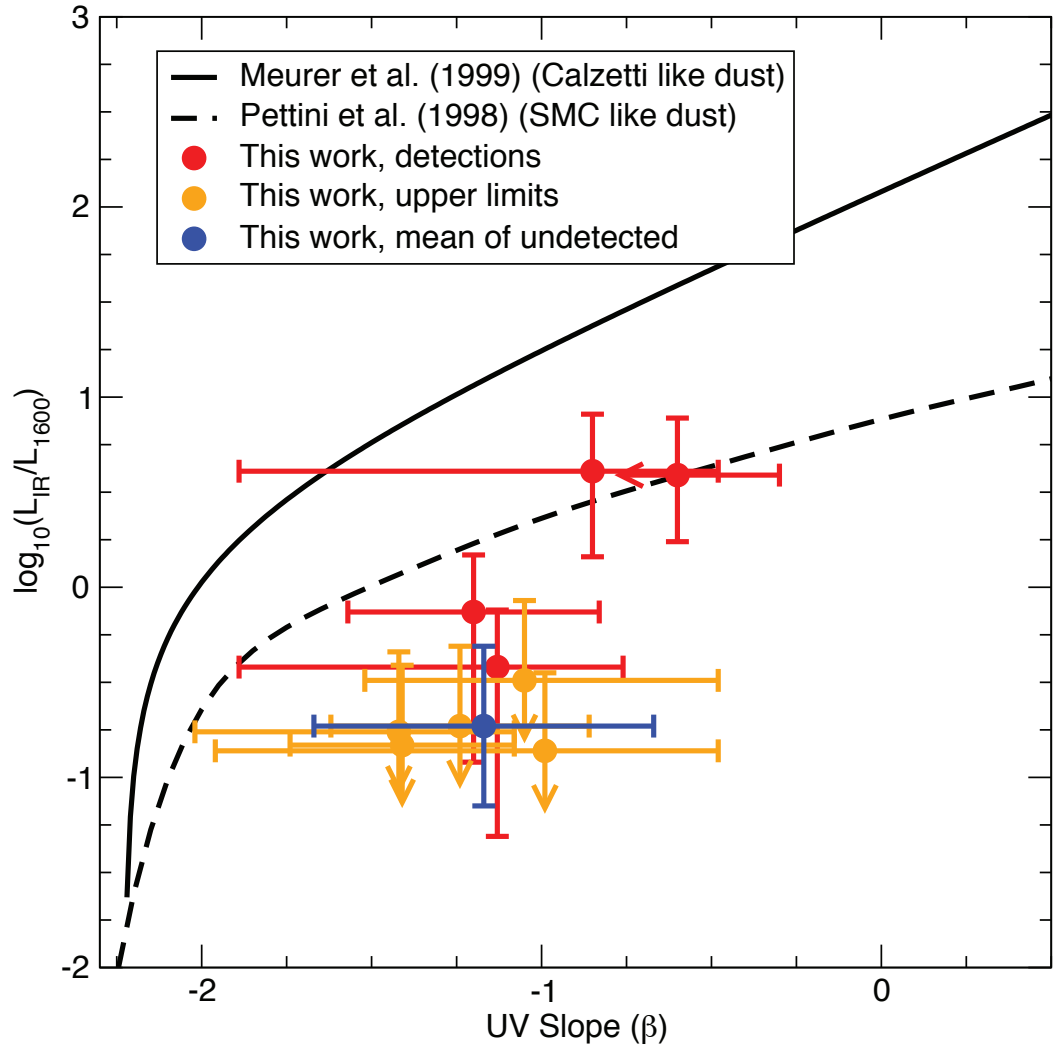

**Figure 2 |** The deficit of infrared emission in our sample is evident in the UV slope (IRX-β) relation when compared with models. Detections are indicated in red, upper limits in orange, and the mean IRX ratio obtained by combining undetected sources is shown in blue. Error bars are 1σ and include standard measurement error and systematic uncertainty added in quadrature. The Meurer[28] relation, consistent with typical galaxies at z<3, is shown as a black solid line, while a model for lower-metallicity SMC like dust model[19] is shown as a dashed line.

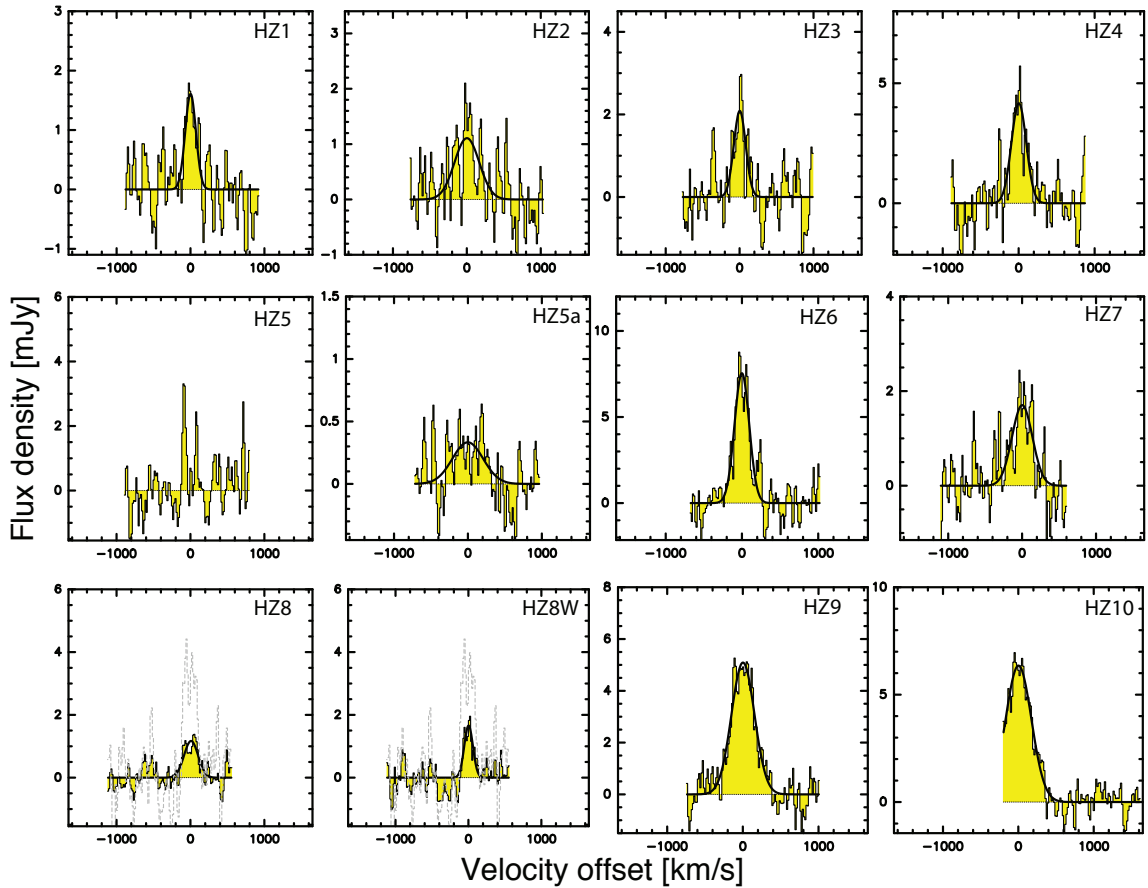

**Figure 3 |** The line velocity profiles of our [CII] measurements are shown our 10 sources and two blind detections HZ5a and HZ8W. All sources but HZ5, a quasar, are clearly detected. For objects HZ8 and HZ8W a dotted line showing that the integrated flux over both sources is also plotted showing they are at similar redshifts.

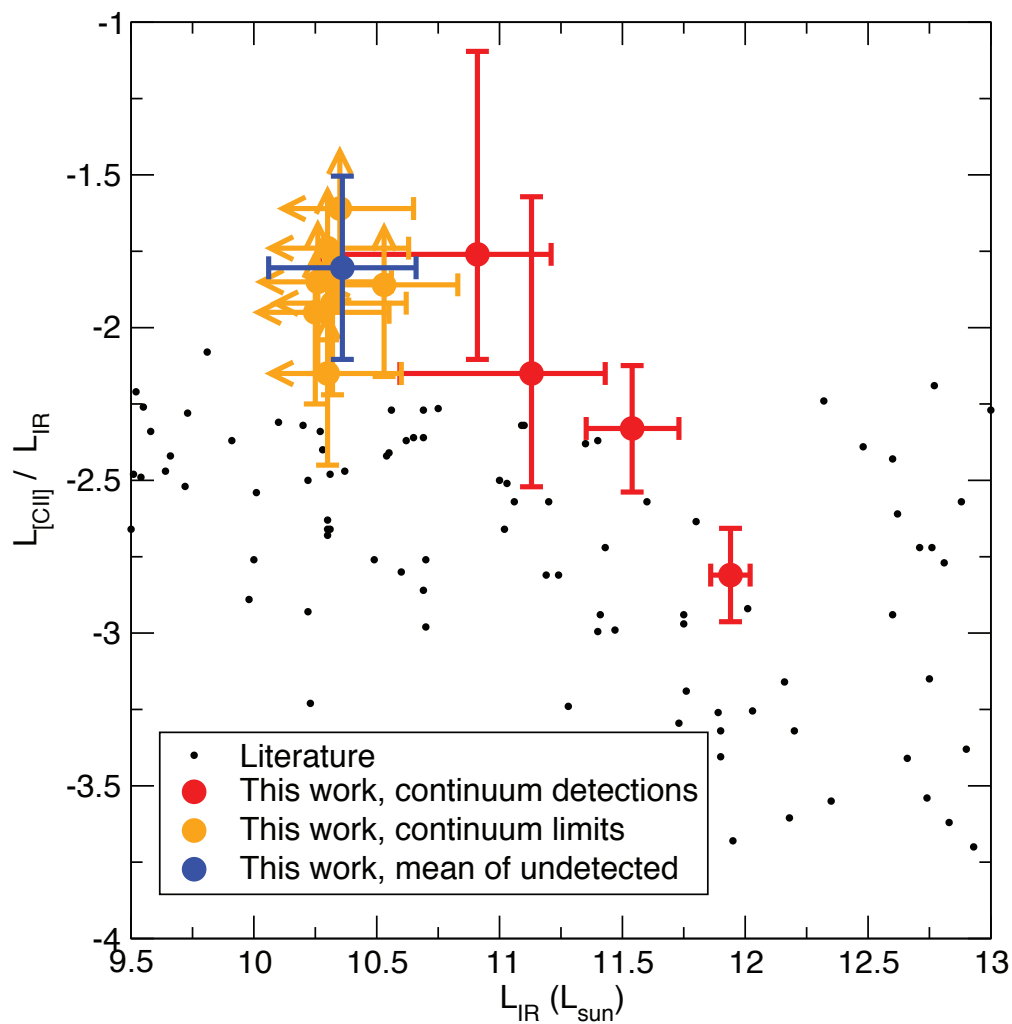

**Figure 4 |** Our sample has a clear excess of [CII] emission relative to FIR emission when compared with normal lower-redshift systems (z=0-3)[6,11,14,29]. Detections are indicated in red, upper limits in orange, and the mean of undetected sources is shown in blue. The high [CII]/FIR ratio we measure indicates a lower dust-to-gas ratio, more diffuse clouds, and a more diffuse UV radiation field than seen in normal z~0 galaxies[11,12,14]. Error bars are 1σ including standard measurement error and systematics added in quadrature.

**Methods**

**1. Cosmology:** Throughout this paper we assume a simplified cosmology of $\Omega_M=0.3$ $\Omega_V=0.7$ $H_o=0.7$, when calculating physical parameters. These values approximates the 2013 Plank[30] and 9-year WMAP[31] values and are commonly assumed in the current literature to make comparisons easier over time as the exact measured values evolve.

**2. Reduction of ALMA data:** The ALMA observatory staff, as part of standard data processing and delivery, performed initial data calibration. The calibrated visibility data was then re-analyzed, performing additional flagging of bad time periods in the data and bad channels. The data were re-imaged using the CASA Briggs flexible weighting of the u-v visibility data with Robust = 1 to create continuum, moment zero, and channel maps. Gaussian fitting of the spectral and spatial data was performed using the CASA viewer and CASA fit tool.

The centroid frequency of these models was compared to the rest frequency of [CII] to measure the redshift, which was compared to the previously determined redshifts from optical spectroscopy. Since these measures were identical, within our precision, the reference frequency of each cube was set to the centroid frequency, allowing the FWHM of the Gaussian to be read in km/s and the line flux in Jy km/s. The [CII] line contribution was subtracted from the measured continuum fluxes.

The measurements derived from the ALMA data are given in Extended Data Table 1 and Gaussian fits done with CASA to the image components are given in Extended Data Tables 2 and 4 for the [CII] and continuum data respectively. The fits were not constrained by the beam size, and so they can be slightly smaller due to the data errors. The errors quoted are those produced by the CASA Gaussian fitting routine and are sensitive to the box size chosen for the fit, especially in cases of low signal to noise. Deconvolved size measurements for the [CII] lines are given in Extended Data Table 3. Only objects 3 and 5a are un-resolved in [CII]. In contrast only object 10 is resolved in the continuum measurement. The baseline configuration used will resolve out spatial scales larger than 5-7" (29-43 kpc) but none of our sources appear to have emission on that large a scale.

**3. Optical and Near-Infrared Data:** Optical and near infrared data were obtained from the COSMOS photometric redshift catalog[32] augmented with additional data from the Spitzer-Large Area Survey with Hyper-Suprime-Cam (SPLASH)[33]. Sizes were taken from the COSMOS Hubble Space Telescope Advanced Camera for Surveys (HST-ACS) weak lensing catalog[15] and are given in Extended Data Table 5. Object 10 falls outside of the area covered by HST-ACS.

**4. Derivation of Physical Parameters:** Since only one FIR data point was available, total infrared (3-1100μm) luminosity was derived using the range of grey body models covering the full range of objects observed at z~2-5[18]. The grey body model parameters, α the blue power-law, β the long wavelength slope, and T the black body temperature, ranged between α=1.5-2.5, β=1.2-2.0 and T=25-45 K. The range of allowed infrared luminosities was determined by scaling these grey bodies to the observed rest frame 158μm flux. This range of parameters resulted in a systematic uncertainty of 0.3 dex in the derived luminosity. This relatively small error is due to the fact that 158μm is on the flat part of the SED close to the emission peak for the range of observed physical parameters. The derived infrared luminosities are given bellow in Extended Data Table 5.

The rest frame ultraviolet properties were determined by fitting a power-law to the rest frame 0.13-0.3μm photometry from the COSMOS and Ultra-Vista surveys. This power-law fit was then used to estimate the UV spectral slope (β) and the luminosity at 0.16μm, where $L_{UV} = \nu_{1600}L\nu_{1600}$. The power-law fits appear to be bias high by the bluest bands in several cases which may be due to the intrinsic SED flatting at short wavelengths [34][35]. To account for this possible systematic we also fit the slope using only the rest-frame 0.15-0.3μm photometry and include the shifts in our measurement error bars. Finally, simulations[2] of observational biases from blending and signal-to-noise indicate an additional ±0.3 systematic error should be included. These estimates are tabulated in Extended Data Table 5.

Stellar masses were derived using data from the Spitzer Large Area Survey with Hyper-Suprime-Cam (SPLASH) in the same way described in previous work[32]. In brief the COSMOS and SPLASH 0.1-5μm photometry was fit with Bruzual and Charlot templates[36] with a Chabrier IMF allowing for a range of emission line strengths and extinction laws. The estimated stellar masses are given in Extended Data Table 5.

Dynamical masses were estimated based on the [CII] velocity dispersion and size estimates in Extended Data Table 3. We used the method outlined in Wang et al.[37] to estimate the dynamical masses. In this approximation $M_{dyn}$ = 1.16x10$^5$ $V_{cir}^2$ D, $V_{cir}$ = 1.763 $\sigma_{[CII]}$/sin(i), and i, the disk inclination angle is approximated by the axis ratio of the object as i = cos$^{-1}$(b/a).

This method yields an approximate answer but is very uncertain because of our low spatial resolution that leads to a large uncertainty in both i, and the object size. The errors quoted in Extended Data Table 5 include the error in the semi-major axis and the velocity dispersion and assumes sin(i)=0.45-1. This range of sin(i) approximates the range of values from a dispersion dominated system to an edge on disk, but could be significantly larger.

We find the dynamical masses are typically a factor of ~3 greater than the stellar masses (See Extended Data Figure 4). This is higher than the factor of 1.2-1.7 found at z~1-3 from similar Hα based dynamical mass estimates[27], but consistent given the very large errors in our measurements.

**5. Robustness of upper limits on dust emission:** The conclusion that the dust properties, and specifically the dust mass and dust to gas ratio, of galaxies are evolving relies on two quantities: the measured deficit of rest-frame 158μm flux, and the models that translate this into a deficit of thermal emission. Here we quantify the robustness of this result.

If we assume a Meurer et al. like IRX-β relation and the mean of our grey body models the predicted mean ALMA flux of the undetected sources is 833-7610μJy (given the uncertainty in β), which is >61σ discrepant from the measured mean of 35±13μJy. Individual sources are discrepant from the Meurer et al. relation by >17-490σ except for objects 9 and 10 which are consistent given the large errors in β. If we assume the maximal range of the systematic error, 45K dust, β=1.5, α=1.2, the predicted mean ALMA flux is 447-4087μJy, which is >32σ discrepant from the measured mean. Finally we consider an extreme model with 100K dust, β=1.5, α=1.2, we predict an ALMA flux of 78-710μJy, which is 3σ discrepant from the measured mean. In this extreme case objects 4, 9, and 10 are consistent with the Meurer et al. relation, and the remaining sources are discrepant at the >1-30σ level.

We also performed a completeness analysis following Casey et al.[38] and find the low dust emission is unlikely due to selection effects in our UV sample. However, we note the result only holds for UV selected samples. The general galaxy population may contain heavily obscured objects that may not be easily found by UV selected surveys.

**6. Quantifying [CII] non-detections in the literature:** Our sample predicts a peak [CII] line flux of ~0.2-0.7 mJy for a $L_*$[39] galaxy with an SFR~6.5 at z~6-7, below the detection limit of most previous studies of z~6-7 objects[6-9].

In addition to sensitivity, we note Ly-α redshifts are systematically offset by ~400 km/s and up to 1200 km/s from the [CII] (and systemic absorption line) redshifts[40], which is a significant fraction of the ALMA band pass.

Finally, caution must be used when interpreting the limits given in the SFR-$L_{[CII]}$ relation because the estimated line luminosity limits scales as the square root of the assumed line width. A 36-100km/s FWHM channel width[6-9] is typically used for these limits, but our measured line widths in Extended Data Table 1 are significantly broader. If we use our measured line widths this would predict [CII] a line limit 0.2-1.0 dex higher than those typically quoted, making them consistent with the SFR-$L_{[CII]}$ relation.

**7. Comparison of Gas and SFR Properties to General Redshift Scaling Relations:** This sample is enhanced in $L_{[CII]}$ relative to $L_{IR}$ compared with lower-redshift samples[6] (Figure 4). These sources have ~1% of their luminosity emitted in the [CII], consistent with low metallicity and extended star forming objects[11,14].

This sample appears to follow the local SFR-$L_{[CII]}$ relation if the SFR is determined by summing the UV and FIR measurements[6] rather than using a dust corrected UV estimate (Extended Data Figure 1). However, if a UV derived dust correction is used the points shift ~0.5-1 dex higher.

Finally we turn to the stellar-mass – SFR or "main-sequence" relation (Extended Data Figure 2)[17]. We find our objects generally agree with this relation, but several objects fall bellow the relation. Specifically objects HZ1, HZ2, and HZ3 below it, but, HZ1 has an unusually low dynamical to stellar mass ratio, indicating the stellar mass may be over-estimated.

**8. Effects of Evolving Dust on the Global Star-Formation History:** Our results indicate the properties of dust and the amount of extinction evolve significantly between z~3, where it is well measured and z~5-6, where we measure it. To estimate the effect on the global Star-Formation history we adopt the luminosity functions and UV slope measurements from the series of papers by Bouwens et al.[2,41], which are largely supported by other work[42]. We then assume the dust correction can be described by the Meurer[28] curve at z<4, the SMC[19] curve at z~5, and the SMC curve scaled to our measured IRX at z~6-8. The results of this analysis are shown in Extended Data Figure 4 and result in ~30-40% less Star-Formation at z~6 than previously thought. For consistency with the literature the Star-Formation rate density is shown assuming a Salpeter IMF, a factor of 0.68 needs to be applied to convert to a Chabrier IMF.

**Method References**

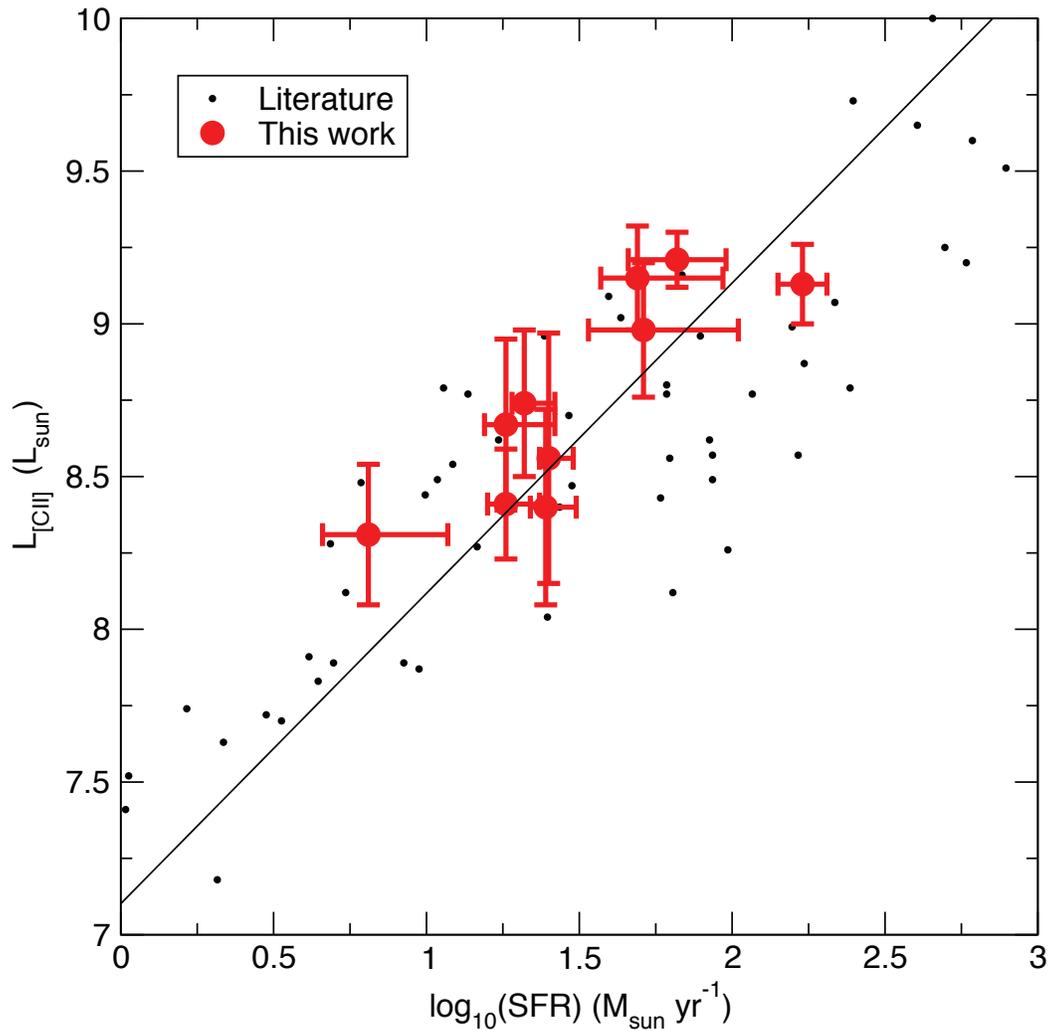

**Extended Data Figure 1 |** Our sample is consistent with the low-z SFR- $L_{[CII]}$ relation from the literature at (z=0-6)[6]. The best fit to this relation from De Looze et al. [43] is indicated with a solid line. The star formation rates are derived using the UV + FIR method and a Chabrier IMF[17]. The dust corrected UV estimates with a Meurer relation typically used at these redshifts would increase the estimated SFR by a factor of ~2-10 (0.5-1 dex), leading to over-estimates of the expected [CII] flux. The $L_{[CII]}$ error bars are 1σ standard measurement error, while the SFR errors are 1σ from a combination of measurement error in $L_{IR}$ and $L_{UV}$ converted to star formation added in quadrature.

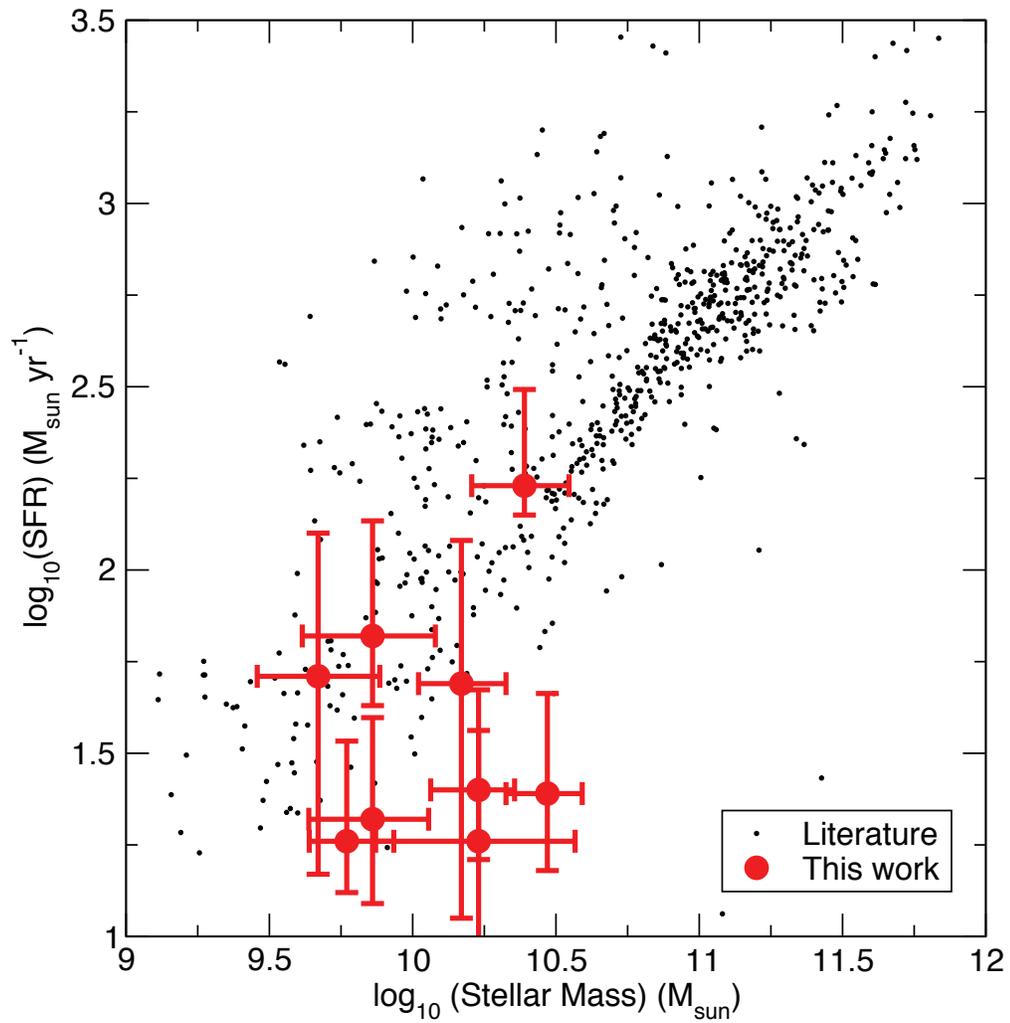

**Extended Data Figure 2 |** Our objects are generally consistent with the stellar mass-SFR or "main sequence" of star forming galaxies at z~5-6[17]. Three objects fall below the relation, HZ1, HZ2, and HZ3, but the stellar mass of HZ1 may be over-estimated (Method 4). Mass errors reflect the 1σ range of masses allowed by the SED model fits including emission line strength as a free parameter. SFR errors are 1σ from a combination of measurement error in $L_{IR}$, systematics in $L_{IR}$, and measurement error in $L_{UV}$ converted to star formation added in quadrature.

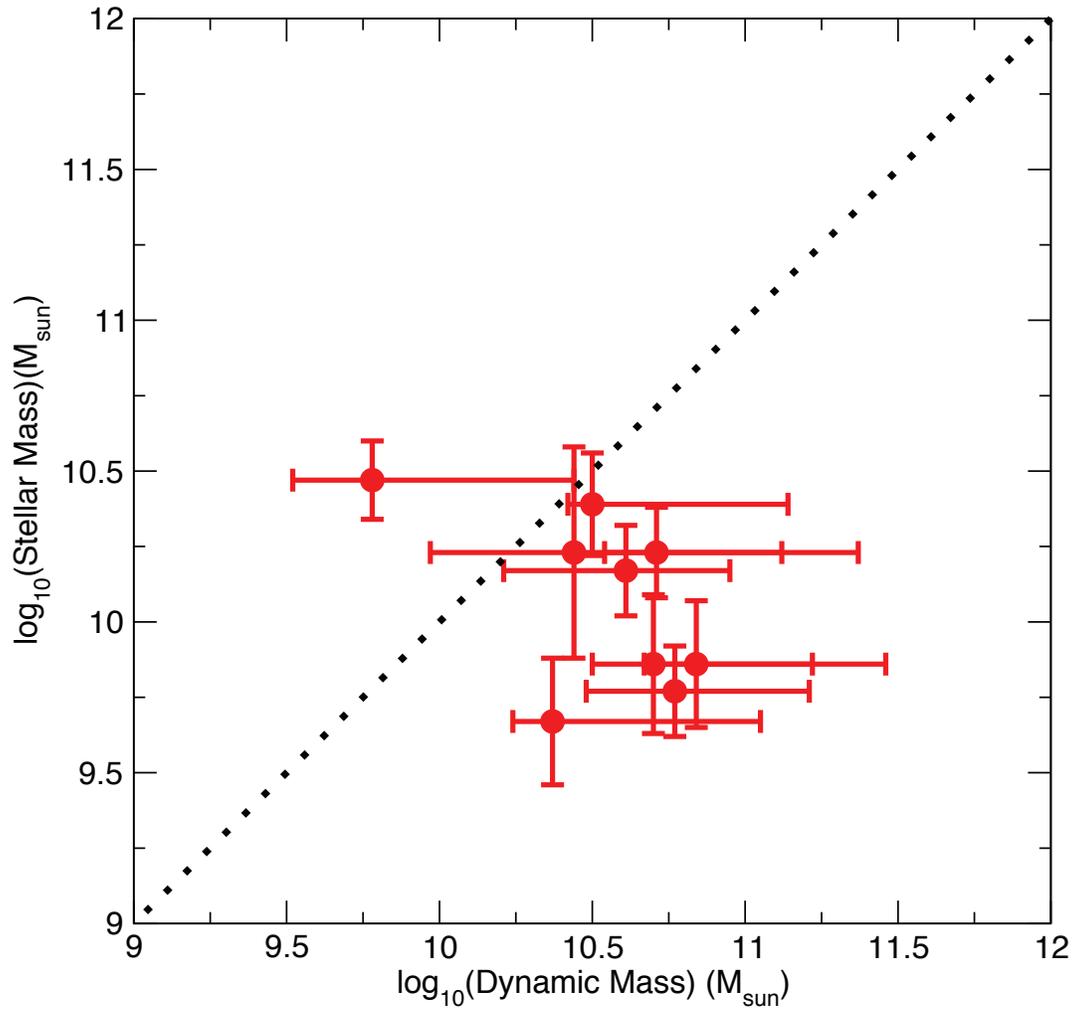

**Extended Data Figure 3 |** A comparison of the dynamical masses estimated from the [CII] line and stellar masses estimated from SED fitting is shown. The average dynamic to stellar mass ratio is ~3 higher than, but consistent with similar measurements at z~1-3[27]. Errors on the dynamical masses reflect the 1σ measurement uncertainty in the size, and velocity dispersion of the sources but the geometry is limited to sin(i)=0.45-1 (Method 4). The errors on the stellar masses reflect the 1σ range of masses allowed by the SED model fits including emission line strength as a free parameter.

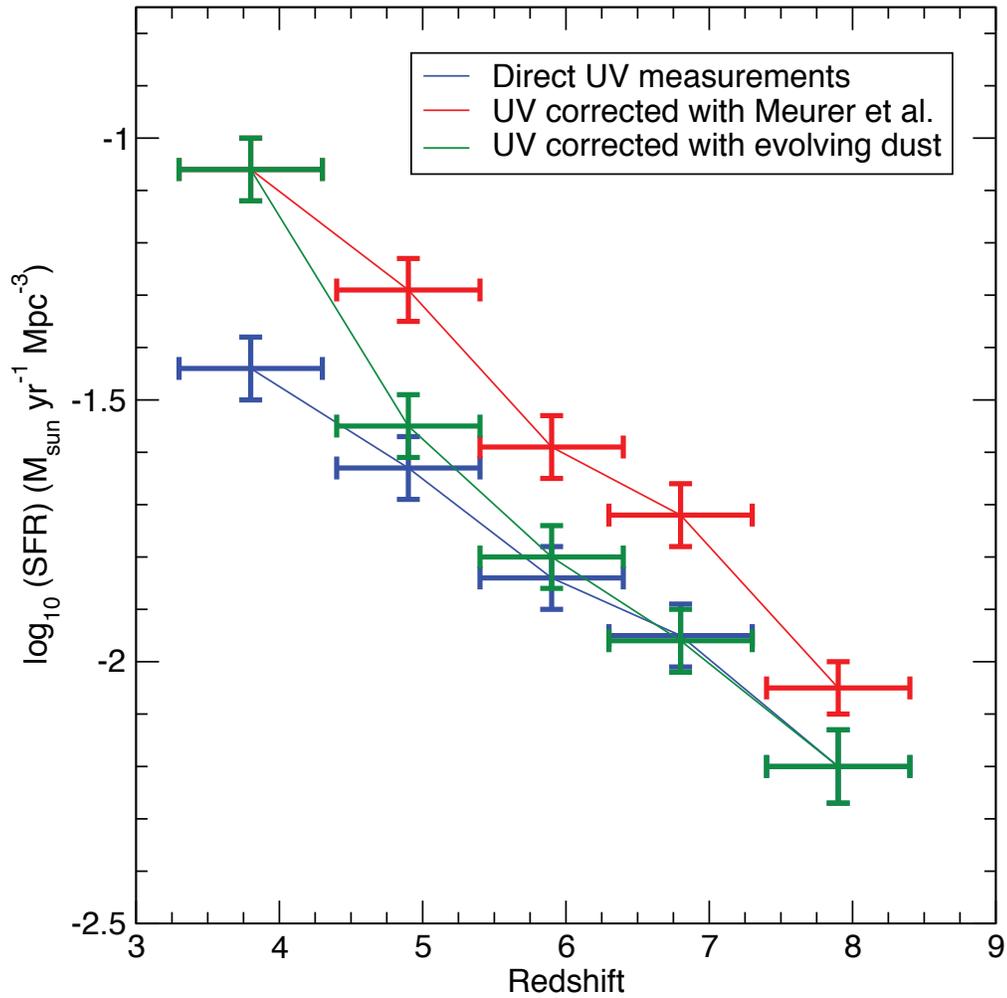

**Extended Data Figure 4 |** The global star formation rate history at z>4 derived from UV measurements[2] are shown for three different assumptions about the dust obscuration in the general galaxy population. No correction for dust are shown in blue, a dust correction assuming the Meurer relation[28] in red, and a correction which linearly evolves in redshift between a Meurer relation at z~4, the SMC like relation at z~5, and our measured value at z~6. Note the evolving dust correction leads to a downward revision by 30-40% at z~6. Redshift error bars reflect the binning of the data, errors in the star formation rate density reflect the 1σ measurement error in the UV luminosity density.

**Extended Data Table 1 | Measured Source Properties**

| ID | RA | DEC | Redshift (UV) II | Selection† | Redshift ([CII]) | Continuum Flux (µJy) | Line Flux (Jy km/s) | σ [CII] (km/s)‡ |
|---|---|---|---|---|---|---|---|---|
| HZ1 | 149.971869 | 2.118175 | 5.690 | LAE/I-LBG | 5.6885±2e-4 | <30 | 0.29±0.04 | 72±11 |
| HZ2 | 150.517074 | 1.928902 | 5.670 | LAE/I-LBG | 5.6697±6e-4 | <29 | 0.42±0.07 | 134±25 |
| HZ3 | 150.039297 | 2.337183 | 5.546 | I-LBG | 5.5416±2e-4 | <51 | 0.56±0.07 | 91±12 |
| HZ4 | 149.618835 | 2.051873 | 5.540 | BBG | 5.5440±2e-4 | 202±47 | 1.14±11 | 108±11 |
| HZ5§ | 150.214993 | 2.582653 | 5.310 | V-LBG | No data | <32 | <0.02 | No data |
| HZ5a | 150.214667 | 2.583136 | No Data | [CII] | 5.3089±1e-4 | <32 | 0.18±0.02 | 20±3 |
| HZ6* | 150.08960 | 2.586411 | 5.290 | V-LBG | 5.2928±1e-4 | 129±36 | 1.80±0.13 | 95±7 |
| HZ7 | 149.876986 | 2.134113 | 5.250 | BBG | 5.2532±4e-4 | <36 | 0.71±0.07 | 162±18 |
| HZ8 | 150.016896 | 2.626614 | 5.148 | R-LBG | 5.1533±2e-4 | <30 | 0.34±0.03 | 107±9 |
| HZ8W | 150.016546 | 2.626731 | No Data | [CII] | 5.1532±1e-4 | <30 | 0.27±0.03 | 63±7 |
| HZ9 | 149.965406 | 2.378379 | 5.548 | R-LBG | 5.5410±1e-4 | 516±42 | 1.95±0.07 | 152±6 |
| HZ10 | 150.247098 | 1.555426 | 5.659 | LAE/I-LBG | 5.6566±2e-4 | 1261±44 | 1.57±0.09 | 127±8 |

\* This source is the same object reported as LBG-1 in Riechers et al. [25]
†LAE=Narrow band selected Ly-alpha emitter, I-LBG = I band dropout Lyman Break Galaxy, R-LBG = R band dropout Lyman Break Galaxy, V-LBG = V band dropout Lyman Break Galaxy, BBG = Balmer Break Galaxy, selected with photo-z and 4.5µm flux, [CII] = [CII] detection in these data.
‡Velocity dispersion, FWHM = 2.35 σ.
§ Chandra detected quasar, shows broad lines in spectra.
II Redshift error is ±4e-3.

**Extended Data Table 2 | Gaussian fits to ALMA [CII] moment zero images**

| ID | Beam Major Axis (") | Beam Minor Axis (") | Beam PA (Degrees) | Source Major Axis (")* | Source Minor Axis (")* | Source PA (Degrees) |
|---|---|---|---|---|---|---|
| HZ1 | 0.82 | 0.51 | 64.2 | 0.93±0.08 | 0.57±0.03 | 69.7±4.2 |
| HZ2 | 0.91 | 0.5 | 64.9 | 1.46±0.37 | 0.69±0.08 | 79.3±5.6 |
| HZ3 | 0.87 | 0.4 | 63.0 | 1.02±0.26 | 0.59±0.1 | 94±12 |
| HZ4 | 0.90 | 0.48 | 59.7 | 1.22±0.15 | 0.61±0.05 | 43.1±4 |
| HZ5 | 0.61 | 0.49 | 65.1 | No Data | No Data | No Data |
| HZ5a | 0.61 | 0.49 | 65.1 | 0.93±0.16 | 0.58±0.07 | 120±10 |
| HZ6 | 0.68 | 0.48 | 65.2 | 1.07±0.14 | 0.83±0.09 | 48±17 |
| HZ7 | 0.71 | 0.49 | 76.8 | 1.19±0.22 | 0.65±0.08 | 72.9±7.9 |
| HZ8 | 0.71 | 0.49 | 61.5 | 1.41±0.21 | 1.01±0.13 | 35±16 |
| HZ8W | 0.71 | 0.49 | 61.5 | 1.05±0.13 | 0.55±0.04 | 75.3±4.2 |
| HZ9 | 0.66 | 0.53 | 75.8 | 0.94±0.07 | 0.66±0.04 | 76.2±6.1 |
| HZ10 | 0.69 | 0.51 | 60.5 | 1.04±0.06 | 0.60±0.02 | 78.7±2.8 |

\* Measured source sizes, not de-convolved.

**Extended Data Table 3 | Deconvolved sizes of [CII] moment zero images.**

| ID | Major Axis (") | Minor Axis (") | PA (Degrees) | Major Axis (kpc) | Minor Axis (kpc) |
|---|---|---|---|---|---|
| HZ1 | 0.45±0.24 | 0.19±0.1 | 84±59 | 2.6±1.4 | 1.1±0.6 |
| HZ2 | 1.17±0.38 | 0.41±0.38 | 86±21 | 6.9±2.2 | 2.4±2.2 |
| HZ3 | <1.02 | <0.59 | ±180 | <6.1 | <3.5 |
| HZ4 | 0.87±0.24 | 0.24±0.16 | 29±18 | 5.2±1.4 | 1.4±1.0 |
| HZ5a | <0.96 | <0.33 | ±180 | <5.8 | <2.0 |
| HZ6 | 0.85±0.21 | 0.65±0.20 | 35±60 | 5.2±1.3 | 4.0±1.22 |
| HZ7 | 0.96±0.30 | 0.42±0.17 | 72±23 | 5.9±1.8 | 2.6±1.0 |
| HZ8 | 1.24±0.28 | 0.84±0.23 | 28±48 | 7.7±1.7 | 5.2±1.4 |
| HZ8W | 0.79±0.19 | 0.19±0.14 | 81±12 | 4.9±1.2 | 1.2±0.9 |
| HZ9 | 0.66±0.09 | 0.39±0.07 | 76±14 | 3.9±0.5 | 2.3±0.4 |
| HZ10 | 0.80±0.09 | 0.28±0.08 | 85.3±5.9 | 4.7±0.5 | 1.7±0.5 |

**Extended Data Table 4 | Gaussian fits to ALMA continuum images** \*

| ID | Beam Major Axis (") | Beam Minor Axis (") | Beam PA (Degrees) | Source Major Axis (") | Source Minor Axis (") | Source PA (Degrees) |
|---|---|---|---|---|---|---|
| HZ4 | 0.87 | 0.48 | 59.2 | 0.82±0.07 | 0.44±0.02 | 56.4±2.9 |
| HZ6 | 0.71 | 0.50 | 65.5 | 0.72±0.06 | 0.45±0.02 | 73.9±4.2 |
| HZ9 | 0.63 | 0.51 | 76.6 | 0.72±0.06 | 0.56±0.04 | 93±11 |
| HZ10 | 0.66 | 0.49 | 61.2 | 1.16±0.11 | 0.59±0.03 | 83.2±3.1 |

\* Only HZ10 is resolved in the continuum with a de-convolved size of major axis = 0.98±0.13", minor axis=0.27±0.12", PA=87.5±5.5.

**Extended Data Table 5 | Derived Physical Properties**

| ID | $L_\odot$ UV[*] | $L_\odot$ IR[†,‡] | SFR ($M_\odot$ yr$^{-1}$) | $L_\odot$ [CII] | Beta[§] | UV r$^{1/2}$ (kpc) | Stellar Mass ($M_\odot$) | Dynamical Mass ($M_\odot$) |
|---|---|---|---|---|---|---|---|---|
| HZ1 | 11.15±0.05 | <10.32 | $24^{+6}_{-3}$ | 8.40±0.32 | -1.41±0.13 | 1.53 | 10.47±0.13 | $9.8^{+1.6}_{-0.8}$ |
| HZ2 | 11.16±0.03 | <10.3 | $25^{+5}_{-2}$ | 8.56±0.41 | $-0.99^{+0.41}_{-0.92}$ | 0.59 | 10.23±0.15 | $10.7^{+1.6}_{-1.0}$ |
| HZ3 | 11.02±0.07 | <10.53 | $18^{+8}_{-3}$ | 8.67±0.28 | -1.05±0.5 | 0.66 | 10.23±0.35 | $<10.4^{+1.6}$ |
| HZ4 | 11.26±0.05 | 11.13±0.54 | $51^{+54}_{-18}$ | 8.98±0.22 | -1.20±0.36 | 0.72 | 9.67±0.21 | $10.4^{+1.6}_{-1.1}$ |
| HZ5[#] | 11.46±0.02 | <10.3 | <3 | <7.2 | $-0.77^{+0.21}_{-0.47}$ | 0.37 | No data | No data |
| HZ5a | <10.3 | <10.3 | <3 | 8.15±0.27 | No data | No data | No data | $<9.0^{+1.8}$ |
| HZ6 | 11.33±0.03 | 10.91±0.64 | $49^{+44}_{-12}$ | 9.15±0.17 | $-1.13^{+0.22}_{-0.71}$ | 3.36 | 10.17±0.15 | $10.6^{+1.6}_{-0.8}$ |
| HZ7 | 11.08±0.04 | <10.35 | $21^{+5}_{-2}$ | 8.74±0.24 | -1.24±0.24 | 0.98 | 9.86±0.21 | $10.8^{+1.5}_{-1.0}$ |
| HZ8 | 11.02±0.06[ǁ] | <10.26 | $18^{+5}_{-2}$ | 8.41±0.18 | $-1.42^{+0.15}_{-0.52}$ | 1.24 | 9.77±0.15 | $10.8^{+1.3}_{-0.9}$ |
| HZ8W | 10.57±0.15[ǁ] | <10.26 | $6^{+5}_{-2}$ | 8.31±0.23 | No data | 1.49 | No data | $9.9^{+1.6}_{-1.2}$ |
| HZ9 | 10.93±0.05 | 11.54±0.19 | $67^{+30}_{-20}$ | 9.21±0.09 | $-0.85^{+0.22}_{-0.95}$[☆] | 0.95 | 9.86±0.23 | $10.7^{+1.3}_{-1.1}$ |
| HZ10 | 11.35±0.06 | 11.94±0.08 | $169^{+32}_{-27}$ | 9.13±0.13 | <-0.6[¶] | No data | 10.39±0.17 | $10.5^{+1.4}_{-1.2}$ |

[*] Monochromatic measured luminosity at rest frame 1600Å with no correction for obscuration. $M_{1600}$ = -2.5*$L_{UV}$ + 5.814
[†] ±0.3 systematic uncertainty in addition to quoted statistical errors.
[‡] 3-1100µm integrated luminosity.
[§] Systematic error is an additional ±0.3 dex based on simulations.
[ǁ] The total UV luminosity for both components was measured as 11.15±0.04 and split between the two components based on the ratio of their observed z' magnitudes.
[¶] Based on the 5σ non-detection in H band.
[#] Known x-ray detected Quasar.
[☆] The low SNR H band point is a significant outlier, error reflects range with that point included and rejected.